\documentclass[pra,nobalancelastpage,twocolumn,superscriptaddress]{revtex4-1}

\usepackage{color,amsthm,amsmath,amsxtra,amsfonts,dsfont,graphicx,bm}
\usepackage[colorlinks=true,linkcolor=blue, citecolor=blue, urlcolor=blue, bookmarks]{hyperref}

\usepackage{amsthm}

\newcommand{\be}{\begin{equation}}
\newcommand{\ee}{\end{equation}}
\newcommand{\bea}{\begin{eqnarray}}
\newcommand{\eea}{\end{eqnarray}}
\newcommand{\bse}{\begin{subequations}}
\newcommand{\ese}{\end{subequations}}

\theoremstyle{plain}

\newcommand{\ket}[1]{\vert#1\rangle}
\newcommand{\bra}[1]{\langle#1\vert}

\begin{document}

\title{Integrability of $1D$ Lindbladians from operator-space fragmentation}

\author{Fabian H.~L. Essler}
\affiliation{The Rudolf Peierls Centre for Theoretical Physics, Oxford
	University, Oxford OX1 3PU, UK}
\author{Lorenzo \surname{Piroli}}
\affiliation{Max-Planck-Institut f{\"{u}}r Quantenoptik,
	Hans-Kopfermann-Str.\ 1, 85748 Garching, Germany}
\affiliation{Munich Center for Quantum Science and Technology, Schellingstra\ss e 4, 80799 M\"unchen, Germany}	

\begin{abstract}
We introduce families of one-dimensional Lindblad equations describing
open many-particle quantum systems that are exactly solvable in the
following sense: $(i)$ the space of operators splits into
exponentially many (in system size) subspaces that are left invariant
under the  dissipative evolution; $(ii)$ the time evolution of the
density matrix on each invariant subspace is described by an
integrable Hamiltonian. The prototypical example is the quantum
version of the asymmetric simple exclusion process (ASEP) which we
analyze in some detail. We show that in each invariant subspace the
dynamics is described in terms of an integrable spin-1/2 XXZ
Heisenberg chain with either open or twisted boundary conditions. We
further demonstrate that Lindbladians featuring integrable operator-space
fragmentation can be found in spin chains with arbitrary local
physical dimension.
\end{abstract}

\maketitle

\section{Introduction}

In many physical situations, a quantum system weakly coupled to an
environment can be described by a Lindblad master
equation~\cite{gorini1976completely,lindblad1976generators}. This
description is valid within a Markovian approximation, where the
internal bath dynamics is assumed to be much faster than that of the
system itself~\cite{breuer2002theory}. The study of many-body Lindblad
equations is generally quite difficult and even in one spatial
dimension the available tools are largely restricted to
perturbative~\cite{li2014perturbative,sieberer2016keldysh} and
sophisticated numerical
approaches~\cite{daley2014quantum,bonnes2014superoperators,Cui2015variational,Werner2016positive,weimer2019simulation}. A
natural question is then whether one can find exactly solvable
Lindblad equations that can give detailed, non-perturbative insights into
representative cases, and provide reliable benchmarks for the
development of approximate methods. This question is also highly
relevant in light of the important role that exactly solvable models
have played in the recent efforts aimed at understand non-equilibrium 
dynamics in isolated quantum systems~\cite{calabrese2016introduction,essler2016quench,caux2016quench,vasseur2016nonequilibrium,calabrese2016quantum,vidmar_generalized_2016}. 

It has been known for some time that certain Lindblad equations
can be cast in the form of imaginary-time Schr\"odinger equations with
non-Hermitian ``Hamiltonians'' that are quadratic in fermionic or
bosonic field operators~\cite{prosen2008third}. This allows one to
obtain exact results on the dynamics of correlation
functions~\cite{eisert2010noise,prosen2010spectral,clark2010exact,Horstmann2013noise,keck2017dissipation,vernier2020mixing}
as well as entanglement-related
quantities~\cite{Somnath2020growth,alba2020spreading}. Furthermore, it 
has also been realized that in some models, not necessarily
integrable, local correlation functions satisfy closed hierarchies of
equations of motion, provided that the Lindbladian satisfies certain
conditions~\cite{eisler2011crossover,vzunkovivc2014closed,caspar2016dissipative,Caspar2016dynamics,Hebenstreit2017Solvable,Foss-Feig2017solvable,klich2019closed}
(although in these cases the full spectrum typically remains out of
reach). Very recently, further progress has been made in this direction, with
the discovery of  Lindbladians that can be mapped onto known
Yang-Baxter integrable \emph{interacting}
models~\cite{Medvedyeva2016Exact,Rowlands2018noisy,Naoyuki2019dissipative,Naoyuki2019dissipativespin,ziolkowska2020yang,nakagawa2020exact,buca2020dissipative,Ribeiro2019integrable,lerma2020trigonometric,yuan_solving_2020}. In particular, the approach put forward in
Refs.~\cite{Medvedyeva2016Exact,ziolkowska2020yang} is based on a map
between Lindblad equations and solvable ``two-leg ladder'' quantum
spin chains. 

Here we highlight a different mechanism that allows us to relate
certain Lindbladians to known integrable models. The systems that we
study are characterized by the fact that the space of operators
``fragments'' into exponentially (in system size) many subspaces that
are left invariant under the dissipative evolution. This is
reminiscent of the fragmentation induced by dipole-conserving
Hamiltonians~\cite{Sala2020Ergodicity,Khemani2020Localization} or confinement~\cite{Yang2020Hilbert},
although in our case this mechanism pertains the space of operators,
and not that of states. For the Lindblad equations studied in this
work, the projection of the Lindbladian onto each invariant subspace
is mapped onto an integrable Hamiltonian, thus allowing us to obtain
the full spectrum analytically. Our construction is carried out for
purely dissipative dynamics, although it is possible to add
certain coherent terms without spoiling integrability. 

We illustrate the above mechanism in a prototypical example: the
quantum version of the celebrated simple asymmetric exclusion process
(ASEP) \cite{SPITZER1970246,Liggett70,DERRIDA199865,Schutz00}. The
classical ASEP is known to be integrable
\cite{Gwa1992six,Gwa1992bethe} and this has made it possible to
obtain a host of exact results, see e.g. 
\cite{Kim1995Bethe,derrida_1999,golinelli2004bethe,de_Gier2005Bethe,Lee_2006,de2006exact,de2008slowest,Simon09,Popkov_2010,deGierEssler11,mallick2011some,crampe2011matrix,Lazarescu_2014,Prolhac_2016}. This
model can be obtained as the averaged dynamics of the so-called 
quantum ASEP~\cite{Jin2020Stochastic}, a stochastic quantum model of
particles hopping with random amplitudes first introduced
in its symmetric form in Ref.~\cite{Bauer2017Stochastic}, see also
Refs.~\cite{Bauer2019Equilibrium,Bernard2019Open,bernard_solution_2020,frassek2020duality}. In this case, we show that the Lindbladian restricted to each invariant
subspace can be mapped onto a XXZ Heisenberg Hamiltonian with
integrable boundary conditions. We also show that a similar
``integrable fragmentation'' can be found in spin chains with
arbitrary local physical dimension.

The rest of this work is organized as follows. We begin in Sec.~\ref{sec:op_fragmentation}, where we introduce the general idea behind the integrable operator-space fragmentation. In Sec.~\ref{sec:asep_lind} we work out in detail the interesting case of the ASEP Lindbladian, discussing both its exact spectrum (Sec.~\ref{sec:spectrum}), and the dissipative dynamics of local observables (Sec.~\ref{sec:dissipative_dyn}). In Sec.~\ref{sec:high_spin_gen} we present higher-spin generalizations, while we report our conclusions in Sec.~\ref{sec:conclusions}. Finally, we consign the most technical aspects of our work to a few appendices.

\section{Operator-space fragmentation}
\label{sec:op_fragmentation}

We consider one-dimensional qudit
systems defined on the Hilbert space $\mathcal{H}=\mathcal{H}_1\otimes
\cdots \otimes \mathcal{H}_L$, with $\mathcal{H}_j\simeq
\mathbb{C}^{d}$. We denote a basis of the local space $\mathcal{H}_j$
by $\ket{\alpha}_{\alpha=1}^d$, and set
$\ket{\boldsymbol{\alpha}}=\ket{\alpha_{1}}_{1}
\ldots\ket{\alpha_{L}}_{L}$. We study the Lindbladian dynamics of the
density matrix $\rho(t)$ of the system defined by $d\rho/dt=-i[H,
\rho]+\mathcal{D}[\rho]$ where 
\be
\mathcal{D}[\rho]\!=\!\sum_{j=1}^{L} \sum_{a=1}^KJ_{a}\!\Big(\!L^{(a)}_{j} \rho L^{(a)\dagger}_{j}-\frac{1}{2}\left\{L^{(a)\dagger}_{j} L^{(a)}_{j}, \rho\right\}\!\Big),
\label{eq:lindblad_eq}
\ee
with $J_a>0$, $K\in\mathbb{N}$, and where periodic boundary conditions are assumed. In the following, we focus on jump operators $L^{(a)}_{j}$ acting on up to two neighboring sites.

We employ a superoperator formalism~\cite{dzhioev2012nonequilibrium},
according to which the density matrix can be viewed as a state in a
$d^{2L}$-dimensional Hilbert space $\mathcal{H}\otimes \mathcal{H}$,
with basis elements
$\ket{\boldsymbol{\alpha}}\ket{\boldsymbol{\beta}}=\ket{\alpha_{1}}_{1}
\ldots\ket{\alpha_{L}}_{L}\ket{\beta_{1}}\!\rangle_{1}
\ldots\ket{\beta_{L}}\!\rangle_{L}$. Given any operator
$\mathcal{O}=\sum_{\boldsymbol\alpha, \boldsymbol\beta}
\mathcal{O}_{\alpha,
  \beta}|\boldsymbol{\alpha}\rangle\langle\boldsymbol{\beta}|$ 
we have the natural mapping $\mathcal{O} \mapsto \ket{\mathcal{O}}=\sum_{\boldsymbol\alpha, \boldsymbol\beta} \mathcal{O}_{\alpha, \beta} \ket{\boldsymbol{\alpha}}\ket{\boldsymbol{\beta}}$. In this formalism, the Lindblad equation can be cast in the form $d|\rho\rangle/d t=\mathcal{L}|\rho\rangle$ where we introduced the Lindbladian $\mathcal{L}=-iH+i\bar{H}+\sum_{j=1}^L\mathcal{D}_j$, and
\begin{align}
\mathcal{D}_j\!=\!\sum_{a=1}^KJ_{a}\!\left[L_{j}^{(a)} \overline{L_{j}^{(a)\dagger}}\!-\!\frac{1}{2}\!\left(\!L_{j}^{(a)\dagger} L_{j}^{(a)}\!+\!\overline{L_{j}^{(a)\dagger} L^{(a)}_{j}}\!\right)\!\right].
\label{eq:liouvillian}
\end{align}
Here we have employed the notation $\mathcal{O}=\mathcal{O}\otimes \openone$, while we defined
$\bar{\mathcal{O}}= \openone\otimes \mathcal{O}^{T}$,  with $\langle\!\bra{\boldsymbol \alpha}\mathcal{O}^{T}\ket{\boldsymbol \beta}\!\rangle=\bra{\boldsymbol\beta} \mathcal{O}\ket{\boldsymbol \alpha}$. 

In the following we assume $H\equiv 0$ (later, we will comment on the possibility of adding simple Hamiltonian terms that do not spoil integrability). For each site $j$, we take $r<d^2$ rank-$1$ projectors $\mathcal{P}^\alpha_{j}$, with $\alpha=1,\ldots,r$, satisfying $\mathcal{P}^{\alpha}_j\mathcal{P}^{\beta}_j=\delta_{\alpha,\beta} \mathcal{P}^{\alpha}_j$, and define $\mathcal{P}_j^0=\openone_{j}-\sum_{\alpha=1}^r \mathcal{P}^{\alpha}_j$. Now, suppose $[\mathcal{L},\mathcal{P}^\alpha_{j}]=0$  for all $\alpha$, $j$. Since
$[\mathcal{P}^{\alpha}_{j},\mathcal{P}^{\beta}_{k}]=0$, the
space $\mathcal{H}\otimes\mathcal{H}$ can be decomposed as a
direct sum of simultaneous eigenspaces $\mathcal{K}_{\boldsymbol{\alpha}}$
of the projection operators $\mathcal{P}^{\alpha}_j$. These subspaces are
labeled by vectors $\boldsymbol{\alpha}=\{\alpha_1\,,\ldots , \alpha_L\}$, specifying that $\mathcal{K}_{\boldsymbol{\alpha}}$ is associated with the projector $\mathcal{P}_1^{\alpha_1}\cdots \mathcal{P}_L^{\alpha_L}$. We note that by construction $\mathcal{K}_{\boldsymbol{\alpha}}$ are left
invariant under the action of the Lindbladian, and their number grows
exponentially with the size $L$. 

Consider a subspace such that $\alpha_{j}\neq 0$ for $j\in\{\ell_1,\ldots
\ell_q\}$, and $\alpha_{j}=0$ otherwise. Physically, the sites labeled by $\ell_j$ can be thought of as ``defects'', separating different regions in the operator-space fragmentation. Indeed, the Lindbladian restricted to
$\mathcal{K}_{\boldsymbol{\alpha}}$ can be expressed as 
\be
\mathcal{L}\big|_{\mathcal{K}_{\boldsymbol{\alpha}}}=\sum_{j=0}^q\mathcal{L}^{\alpha_{\ell_j},\alpha_{\ell_{j+1}}}_{[\ell_{j},\ell_{j+1}]}\,,
\label{eq:decomposition}
\ee
with $\ell_{0}=\ell_q$ and $\ell_{q+1}=\ell_1$. Here, we have defined
\begin{align}
\!\!\mathcal{L}^{\alpha,\beta}_{[m,n]}&\!=\!\mathcal{P}^{\alpha}_{m}\mathcal{P}^0_{m+1}\mathcal{D}_{m}\mathcal{P}^{\alpha}_{m}\mathcal{P}^{0}_{m+1}
	\!+\!\mathcal{P}^0_{n-1}\mathcal{P}^{\beta}_{n}\mathcal{D}_{n-1}\mathcal{P}^{0}_{n-1}\mathcal{P}^{\beta}_{n}\nonumber\\
	&+\sum_{j=m+1}^{n-1}\mathcal{P}^0_{j,j+1}\mathcal{D}_{j}\mathcal{P}^0_{j,j+1}\,,\quad |m-n|>1\,,
	\label{eq:projected_liouvillian}
\end{align}
and $\mathcal{L}^{\alpha,\beta}_{[m,m+1]}=\mathcal{P}^{\alpha}_{m}\mathcal{P}^{\beta}_{m+1}D_m\mathcal{P}^{\alpha}_{m}\mathcal{P}^{\beta}_{m+1}$, where $\mathcal{P}^0_{j,j+1}=\mathcal{P}^0_{j}\mathcal{P}^0_{j+1}$, and $\mathcal{D}_j$ is given in Eq.~\eqref{eq:liouvillian}. Our observation is that there exist simple families of jump operators $L_{j}^{(a)}$ such that the action of $\mathcal{L}^{\alpha,\beta}_{[m,n]}$ coincides with that  of an integrable Hamiltonian with appropriate boundary terms. When this happens, the full spectrum of  the Lindbladian can be obtained analytically, by analyzing each invariant sector individually. Importantly, we stress that $\mathcal{L}^{\alpha,\beta}_{[m,n]}$ acts on a space where the effective local dimension is not necessarily the square of an integer number, allowing for a mapping to different integrable systems with respect to those found in Refs.~\cite{Medvedyeva2016Exact,ziolkowska2020yang}.

\section{The ASEP Lindbladian}
\label{sec:asep_lind}
 
As a concrete example, we consider the fully dissipative Lindbladian defined by Eq.~\eqref{eq:liouvillian}, with $d=K=2$, and 
\be
L_{j}^{(1)}=\sigma_{j}^{+} \sigma_{j+1}^{-}, \quad L_{j}^{(2)}=\sigma_{j}^{-} \sigma_{j+1}^{+}\,.
\label{eq:ASEP_jump}
\ee 
Here $\sigma^\pm=(\sigma^x\pm i\sigma^y)/2$, and $\sigma^\alpha$ are
the Pauli matrices. This Lindbladian admits an equivalent
representation in terms of spinless fermionic operators via a standard
Jordan-Wigner mapping~\cite{wigner1928paulische}, and in this
formulation it can be obtained after averaging over the bath degrees
of freedom in the ``quantum ASEP'' studied in Refs.~
\cite{Jin2020Stochastic,Bauer2017Stochastic,Bauer2019Equilibrium,Bernard2019Open,bernard_solution_2020}. It
was already pointed out in these works that the above Lindblad
equation reduces to the classical ASEP in the subspace of 
density matrices that are diagonal in the position basis introduced
above~ \footnote{For this reason, in the following we will denote the
  Lindbladian restricted to this subspace by $\mathcal{L}_{\mathrm{ASEP}}$.}. When combined with the known
mapping between the classical ASEP and the quantum XXZ Heisenberg 
chain~\cite{Gwa1992six,Gwa1992bethe}
this gives us the integrability of the Lindbladian restricted to this
subspace. In the following, we establish a stronger statement, namely
we show that its full spectrum can be obtained
analytically, as a result of the operator-space fragmentation.  

It is convenient to order the basis elements generating the local
Hilbert space $\mathcal{H}_j\otimes \mathcal{H}_j$ as
$\ket{e_1}=\ket{1}\otimes \ket{1}\!\rangle$, $\ket{e_2}=\ket{2}\otimes
\ket{1}\!\rangle$, $\ket{e_3}=\ket{1}\otimes \ket{2}\!\rangle$,
$\ket{e_4}=\ket{2}\otimes \ket{2}\!\rangle$ and define, accordingly, the
operators $E^{\alpha,\beta}$ by their action
$E^{\alpha,\beta}\ket{e_{\gamma}}=\delta_{\beta,\gamma}\ket{e_{\alpha}}$. With
these notations, it is straightforward to see that $\mathcal{L}$
commutes with $\mathcal{P}^{1}_j=E_j^{2,2}$ and $\mathcal{P}^{2}_j=E_j^{3,3}$, so that $\mathcal{P}_j^0$ is the projector onto
the subspace spanned by $\ket{\alpha}_j\ket{\alpha}\rangle_j$

Following the general discussion given before, we focus on the
Lindbladian restricted to different subspaces
$\mathcal{K}_{\boldsymbol{\alpha}}$. Let us first consider the case where
$\alpha_j=0$, for all $j$. Its restriction to
$\mathcal{K}_{\boldsymbol{\alpha}}$ acts on the tensor product of
$2$-dimensional local spaces, which are spanned by the states
$\ket{e_1}_j$, $\ket{e_4}_j$. Choosing this basis for each site, we show in Appendix~\ref{sec:quantum_asep} that
the restricted Lindbladian can be rewritten as
\begin{align} 
\mathcal{L}_{\mathrm{ASEP}}&=\sum_{j=1}^L\Big[ J_{1} \sigma_{j}^{+}
\sigma_{j+1}^{-}+J_{2} \sigma_{j}^{-} \sigma_{j+1}^{+}\nonumber\\
&\qquad\qquad+\frac{J_{1}+J_2}{4}\big(\sigma^z_j\sigma^z_{j+1}-1\big)\Big].
\label{eq:ASEP_L_periodic_main}
\end{align}
This is precisely the evolution operator of the \emph{classical} ASEP
with periodic boundary conditions, which is known to be integrable
\cite{Gwa1992six,Gwa1992bethe}. Applying a local similarity
transformation $S=\prod_j S_j$, with $S={\rm diag}(1,\sqrt{J_1/J_2})$,
$\mathcal{L}_{\mathrm{ASEP}}$ can be mapped to a spin-1/2 Heisenberg
Hamiltonian with twisted boundary conditions, cf. Appendix~\ref{sec:quantum_asep}. The
Lindbladian (\ref{eq:ASEP_L_periodic_main}) describes time evolution of
quantum states that are by construction \emph{unentangled}.

When $\alpha_j\neq 0$  for some $j\in\{\ell_1,\ldots,\ell_q \}$, instead, the Lindbladian can be decomposed
as in Eq.~\eqref{eq:decomposition}. If $|\ell_j-\ell_k|\leq 3$, the projected Lindbladian is simply a constant. Otherwise, we find
\be
\mathcal{L}^{\alpha,\beta}_{[m,n]}=\openone_{m}\otimes\tilde{\mathcal{L}}_{[m+1,n-1]}\otimes \openone_{n}\,, 
\label{eq:factorization}
\ee
where $S^{-1}\tilde{\mathcal{L}}_{[m,n]}S=
\mathcal{J}(-H^{\rm XXZ}_{[m,n]}-2\Delta)$ and
\be
H^{XXZ}_{[m,n]}\!=\!-\!\sum_{j=m}^{n-1}\! \left[ \sigma_{j}^{x} \sigma_{j+1}^{x}\!+\!\sigma_{j}^{y} \sigma_{j+1}^{y}\!+\!\Delta\!\left(\sigma_{j}^{z} \sigma_{j+1}^{z}\!-\!1\right) \right]\,.
\label{eq:XXZ_hamiltonian_open}
\ee
Here we have defined $\mathcal{J}=\sqrt{J_1J_2}/2$, $Q=\sqrt{J_1/J_2}$ and
$\Delta=(Q+Q^{-1})/2$. The Lindbladian $\mathcal{L}^{\alpha,\beta}_{[m,n]}$ are
required in order to describe time evolution of quantum states with
non-vanishing entanglement. The operators
\eqref{eq:ASEP_L_periodic_main}, \eqref{eq:XXZ_hamiltonian_open} are
integrable, corresponding to the XXZ Heisenberg chain with diagonal
twisted~\cite{alcaraz1990spin,albertini1996phase} and
open~\cite{alcaraz1987surface} boundary conditions, respectively. \\ [-0.15cm]

Before leaving this section, we sketch how a coherent term can be added to the Lindbladian without spoiling integrability. It is easy to see that, for the quantum ASEP, any Hamiltonian whose action preserves the invariant subspaces (namely, that is written entirely in terms of Pauli matrices $\sigma^z_j$) also preserves integrability. The most general nearest-neighbor Hamiltonian of this form is 
\be
H=\sum_{j=1}^L J^{\prime} \sigma_{j}^{z} \sigma_{j+1}^{z}+h \sigma_{j}^{z}\,.
\label{eq:hamiltonian}
\ee
In Appendix~\ref{sec:quantum_asep}, we show that the projection of the corresponding term $i(H-\bar{H})$ onto the different invariant subspaces results in diagonal boundary terms, thus preserving integrability. 

\subsection{The exact spectrum} 
\label{sec:spectrum}

We now study the full spectrum of the Lindbladian defined by the jump operators~\eqref{eq:ASEP_jump}. Let us  first consider the case where $\alpha_j=0$ for all $j$, corresponding to Eq.~\eqref{eq:ASEP_L_periodic_main}. It is easy to see that
the ground-state energy is $E_{\rm GS}=0$, with degeneracy $L+1$. The
associated eigenspace is spanned by $\ket{{\rm gs}_{M}}$, $M=
0,1,\ldots L$, where $\ket{{\rm gs}_M}$ is the equal-weight
superposition of all spin configurations in the sector of $M$ ``down
spins''. We note that $\ket{{\rm gs}_{M}}$ corresponds to the
projection of the non-equilibrium steady state $\rho_{\rm
  NESS}=\openone/2^L$ onto a given magnetization sector. It is worth
mentioning that, although $\rho_{\rm NESS}$ is formally an
infinite-temperature state, it features a non-vanishing expectation
value of the spin current, whose density $J^z_{j}$ is defined by ${\rm
  tr}(\dot\rho(t)\sigma_j^z)=-{\rm tr}[\rho(t) (J^z_{j+1}-J^z_{j})]$.  

One can study the full spectrum of  $\mathcal{L}_{\rm ASEP}$ (and the associated eigenstates) via a standard application of the Bethe Ansatz formalism. In fact, a detailed analysis has been already reported in Ref.~\cite{Gwa1992bethe}, in the context of the classical ASEP, where it was shown that the Hamiltonian features a spectral gap vanishing as $L^{-{3/2}}$.  In the following, we sketch how an exact analysis can be carried out also for the spectrum of all other projected Lindbladians.

Let us then consider the case where  $\alpha_j\neq 0$ for some $j$. We immediately see that, since $H^{\rm XXZ}_{[m,n]}\geq 0$, the eigenvalues $\lambda_k$ of $ \tilde{\mathcal{L}}_{[m,n]}$ in Eq.~\eqref{eq:factorization} are bounded by  
\be
\lambda_k\leq -(J_1+J_2)/2\,,
\label{eq:bound}
\ee
where the bound is strict, since the smallest eigenvalue of $H^{\rm XXZ}_{[m,n]}$ is $E^0_{\rm GS}=0$ (corresponding to the two saturated ferromagnetic states). This means that the long-time limit of the Lindbladian evolution is dominated by the subspace of locally-diagonal density matrices, containing all the ``soft modes'' of the dynamics. Once again, the full spectrum of $H^{\rm XXZ}_{[m,n]}$  can be obtained using the Bethe Ansatz formalism~\cite{alcaraz1987surface}. According to the latter, each eigenstate is associated with a set of quasi-momenta $\{\lambda_j\}_{j=1}^M$, satisfying the set of quantization conditions
\be
\left[S_1(\lambda_j)\right]^{2 \ell}f\left(\lambda_{j}\right)=\prod_{k \neq j} S_2\left(\lambda_{j}-\lambda_{k}\right) S_2\left(\lambda_{j}+\lambda_{k}\right)\,,
\ee
where $\ell=n-m+1$ is the length of the open chain, and
\be
f\!\left(x\right)\!=\!\frac{\cos^{2} \left(x-i \eta/2\right)}{\cos^{2} \left(x+i \eta/2\right)},\ S_n\left(x\right)\!=\!\frac{\sin \left(x+i n\eta/2\right)}{\sin \left(x-i n\eta/2\right)}\,,
\ee
with $\eta=\operatorname{arccosh}(\Delta)$. The associated energy eigenvalue is then $E\left[\{\lambda_j\}_{j=1}^M\right]=\sum_{j=1}^{M}\varepsilon(\lambda_j)$, where $\varepsilon(x)=4\sinh^2(\eta)/[\cosh\eta-\cos(2x)]$.

One can explicitly identify the eigenstate associated with the
lowest-energy eigenvalue $E^{M}_{\rm GS}$ in each magnetization
sector, by following the derivation in Ref.~\cite{albertini_xxz_1995},
where the analogous problem for periodic boundary conditions was
considered. We report this analysis for $M=\ell/2$ ``down spins''
(assuming $\ell$ to be even). By numerically solving the Bethe Ansatz
equations for small systems we find that the smallest eigenvalue in
this sector corresponds to a maximal string with real part $\pi/2$,
namely to a solution of the Bethe equations of the form
$\lambda_{j}=\frac{\pi}{2}+i(2 j-1) \frac{\gamma}{2}+\delta_{j}, \quad
j=1, \ldots, M$, where $\delta_j$ are deviations from a ``perfect 
string''. The ground state is doubly degenerate in the thermodynamic
limit, and at finite size the exponentially close ground states
correspond to different values of the deviations $\delta_j$. Assuming
$\delta_j$ to vanish as $\ell\to \infty$, and repeating the steps
reported in Ref.~\cite{albertini_xxz_1995}, one finds   
\be
E^{\ell/2}_{\rm GS}=2\sqrt{\Delta^2-1}+O(e^{-a \ell})\,,
\ee
where $a>0$ is a constant (which depends on $\Delta$).  Note
that, as in the case of periodic boundary conditions, low-lying
excitations are separated by a finite gap from $E^{M}_{\rm GS}$. A
careful analysis of the deviations $\delta_j$ for large systems, and
the identification of solutions corresponding to higher ``excited
states'' are of interest in particular in light of the complications
that are known to arise for $\Delta=1$
\cite{Sutherland95,DharShastry00}, but are beyond the scope of this work.

Due to Eq.~\eqref{eq:bound}, we see that the invariant subspaces are
organized in hierarchies of decreasing energy. Indeed, given $\boldsymbol{\alpha}$, we can identify the subset $\{\alpha_{\ell_j}\}$ such that $\alpha_{\ell_j}\neq 0$, and as long as $|\ell_j-\ell_{j+1}|>1$, the Lindbladian restricted to the corresponding invariant subspace lowers  the ground-state energy by an amount $-(J_1+J_2)/2$. If $|\ell_j-\ell_{j+1}|=1$ for all $j$, then we have a completely fragmented
space ($\alpha_j\neq 0$ for all $j$), and a dedicated analysis is needed. As we show in Appendix~\ref{sec:quantum_asep}, in this case, any product state is fixed under the Lindbladian evolution, and thus
corresponds to an eigenstate of the Lindbladian. The associated
eigenvalue is always bounded by $-(J_1+J_2)/2$, except for
the two special states $\ket{\boldsymbol{e_2}}=\ket{e_2}^{\otimes L}$
and $\ket{\boldsymbol{e_3}}=\ket{e_3}^{\otimes L}$ that are
annihilated by $\mathcal{L}$, making its ground-state degeneracy
$L+3$.\\ [-0.15cm]

\subsection{The dissipative dynamics}
\label{sec:dissipative_dyn}

Any initial density matrix can be decomposed as
$|\rho(0)\rangle=\sum_{\boldsymbol{\alpha}} \rho_{\boldsymbol{\alpha}
  \boldsymbol{\alpha}}|\boldsymbol{\alpha}\rangle|\boldsymbol{\alpha}\rangle\!\rangle+\sum_{\boldsymbol{\beta}
  \neq \boldsymbol{\alpha}} \rho_{\boldsymbol{\alpha}
  \boldsymbol{\beta}}|\boldsymbol{\alpha}\rangle|\boldsymbol{\beta}\rangle\!\rangle$. Due
to Eq.~\eqref{eq:bound}, we have that, up to terms that are
exponentially small in time,
\begin{equation}
\!\!|\rho(t)\rangle\!=\! \sum_{\boldsymbol{\alpha}}
\rho_{\boldsymbol{\alpha} \boldsymbol{\alpha}}
e^{\mathcal{L}_{\mathrm{ASEP}}
  t}|\boldsymbol{\alpha}\rangle|\boldsymbol{\alpha}\rangle\!\rangle
+{\cal O}(e^{-(J_1+J_2)t/2})\ ,
\end{equation}
and the late-time behavior is therefore fully determined by the
classical ASEP. Here we have neglected the subspace associated with
the fixed points $\ket{\boldsymbol{e_2}}$, $\ket{\boldsymbol{e_3}}$,
which is expected to yield contributions that are exponentially small
in the system size $L$. On the other hand, for ``quantum'' operators
that have no analogues in the classical ASEP the dynamics takes place
entirely in subspaces that are orthogonal to that spanned by
$\ket{\boldsymbol{\alpha}}\ket{\boldsymbol{\alpha}}\!\rangle$. In
order to illustrate this, we consider the transverse spin-spin
correlation function 
\be
S_{1, \ell}^{+-}(t)=\operatorname{Tr}\left[\rho(t) \sigma_{1}^{+} \sigma_{\ell}^{-}\right]\,.
\ee
In the superoperator formalism, this can be rewritten as $S_{1, \ell}^{+-}(t) =\left\langle \boldsymbol{\phi}\left|E_{1}^{12} E_{\ell}^{43}\right| \rho(t)\right\rangle$, where $\bra{\boldsymbol{\phi}}=(\bra{e_1}+\bra{e_4})^{\otimes L}$.
This shows that this two-point function is only sensitive to the
invariant subspace corresponding to $\alpha_{1}=1$, $\alpha_{\ell}=2$ and
$\alpha_{j}=0$ otherwise. Assuming for simplicity $\ell\geq 3$,  this leads to
\begin{align}
S_{1, \ell}^{+-}(t)&=\sum_{\boldsymbol{a},\boldsymbol{b}}  \rho_0\left(\boldsymbol{a}, \boldsymbol{b}\right)\langle \phi^{[2, \ell-1] }|e^{\tilde{\mathcal{L}}_{[2, \ell-1]} t}| \boldsymbol{a}\rangle \nonumber\\
& \times \langle \phi^{[\ell+1,L]}|e^{\tilde{\mathcal{L}}_{[\ell+1, L]}
  t}| \boldsymbol{b}\rangle\,,
\label{S1spec}
\end{align}
where we have defined $\langle \phi^{[m, n]}|=\otimes_{j=m}^n
(\bra{e_1}_{j}+\bra{e_4}_{j})$, $\ket{\boldsymbol{a}}=\ket{e_{a_1}}\ket{e_{a_2}} \cdots
\ket{e_{a_{\ell-2}}}$, $\ket{\boldsymbol{b}}
=\ket{e_{b_1}} \cdots \ket{e_{b_{L-\ell}}}$,
for $a_j,b_j=1,4$ and 
$\rho_0(\boldsymbol{a},\boldsymbol{b})=\langle
\boldsymbol{a}^\prime |\langle
\boldsymbol{b}^\prime|\rho(0)\rangle$, where $\ket{\boldsymbol{a}^\prime}=\ket{e_{2}}\ket{\boldsymbol{a}}$,
$\ket{\boldsymbol{b}^\prime} =\ket{e_{3}}\ket{\boldsymbol{b}}$. Eq.~\eqref{S1spec} reduces the problem of calculating the transverse spin-spin
correlator to computing $\langle \phi^{[m,n]
}|e^{\tilde{\mathcal{L}}_{[m,n]} t}| \boldsymbol{a}\rangle$. This
is still non-trivial because $\bra{ \phi^{[m,n]  }}$ is not an
eigenstate of $\tilde{\mathcal{L}}_{[m,n]}$, due to the boundary
terms. A simplification occurs in the isotropic limit $J_1=J_2$, where
$\bra{\phi^{[m,n] }}$ becomes a left eigenstate of
$\tilde{\mathcal{L}}_{[m,n]}$ with eigenvalue $-J$. Hence we have
\be
S_{1,\ell}^{+-}(t)=e^{-2 J t} S_{1,\ell}^{+-}(0)\,.
\ee
We note that this result could be obtained in a direct way from the
equations of motion for $\sigma^+_1\sigma^-_{\ell}$, which for
$J_1=J_2$ become linear. For $J_1\neq J_2$ the calculation of
$\langle \phi^{[m,n] }|e^{\tilde{\mathcal{L}}_{[m,n]} t}|
\boldsymbol{a}\rangle$ is an open question, but we note that
similar quantities have been recently determined in problems of real-
and imaginary-time evolution from simple product
states~\cite{pozsgay2013dynamical,piroli2017quantum,piroli2018non}.

\section{Higher spin generalizations}
\label{sec:high_spin_gen}

 Lindbladians featuring similar types of fragmentation can be constructed also for higher local dimension. We consider a Hilbert space $\mathcal{H}=\mathcal{H}_1\otimes \cdots \otimes \mathcal{H}_L$, with $\mathcal{H}_j\simeq \mathbb{C}^{N}$ and a purely dissipative Lindblad equation of the form $\dot{\rho}=-\sum_j\mathcal{D}_j(\rho)$, with $\mathcal{D}_j$ as in Eq.~\eqref{eq:liouvillian}. We choose $K=N^2$, $J_a=1$, and jump operators~\footnote{We note that related Lindbladians have been recently considered in Ref.~\cite{frassek2020duality}, where they were shown to describe the average quantum SSEP in a $n$-replica space.}
\be
L^{(\alpha,\beta)}_{j}=E^{\alpha,\beta}_{j}E^{\beta,\alpha}_{j+1}\,,\quad \alpha,\beta=1\,,\ldots \,, N\,.
\ee
Here, $E^{\alpha,\beta}$ is the operator with matrix elements
$(E^{\alpha,\beta})_{\alpha^\prime,\beta^\prime}=\delta_{\alpha,\alpha^\prime}\delta_{\beta,\beta^\prime}$.
In the superoperator formalism, it is readily seen that $\mathcal{L}=\mathcal{L}^\prime-L\openone $, where $\mathcal{L}^\prime$ commutes with the $d(d-1)$ rank-$1$ projectors
$\mathcal{P}^{(\alpha,\beta)}_j= E^{\alpha,\alpha}_j\otimes
E^{\beta,\beta}_j$ for $\alpha\neq \beta$, $\alpha,\beta=1,\ldots, d$, so that $\mathcal{P}^{0}_j=\openone_j-\sum_{\alpha\neq\beta}\mathcal{P}^{(\alpha,\beta)}_j$ projects onto the subspace spanned by the ``diagonal'' states  $\ket{e_\alpha}_j=\ket{\alpha}_j\otimes \ket{\alpha}\!\rangle_j$, $\alpha=1\,,\ldots \,, N$. 
This implies that the Hilbert space $\mathcal{H}\otimes \mathcal{H}$
splits into exponentially many invariant subspaces
$\mathcal{K}_{\boldsymbol{\alpha}}$.  We start by analyzing the one corresponding to
$\alpha_j=0$ for all $j$. In this case the local physical dimension is
$N$ and choosing the local diagonal basis introduced above, the projection of $\mathcal{L}^\prime$ can be rewritten as
\be
\mathcal{L}^\prime\big|_{\boldsymbol{\alpha}=0}=\sum_{j=1}^{L}\Pi_{j, j+1}\,,
\label{eq:final_lind_N}
\ee
where $\Pi_{j,j+1}$ is the operator swapping neighboring sites, cf. Appendix~\ref{sec:higher_spin}. We see that $\mathcal{L}^\prime\big|_{\boldsymbol{\alpha}=0}$ is indeed integrable, as its action coincides with that of the $SU(N)$-invariant Sutherland model, first solved in Ref.~\cite{Sutherland1975model}. Next, we consider the generic case where $\alpha_j\neq 0$  for some $j\in\{\ell_1,\ldots,\ell_q \}$. If $|\ell_j-\ell_k|\leq 3$, the projection of $\mathcal{L}^\prime$ vanishes; otherwise, we have the factorization~\eqref{eq:factorization} with
\be
\tilde{\mathcal{L}}_{[m,n]}=\sum_{j=m}^{n-1}\Pi_{j, j+1}\,.
\label{eq:open_suN}
\ee
This is the $SU(N)$-invariant chain with diagonal boundary conditions, which is once again integrable~\cite{de_vega1994exact,de_vega1994exactII,doikou1998bulk} and,  as a result, the full spectrum can be analyzed analytically via the Bethe Ansatz. We mention that, although here we have focused on $J_a=1$ for all $a$, we expect that one could also choose $J_a$ such that each projected Lindbladian is mapped onto an integrable deformation of the isotropic Sutherland model.

\section{Conclusions} 
\label{sec:conclusions}

We have studied families of Lindbladians that can be mapped onto known Yang-Baxter integrable models. These systems are characterized by a fragmentation of the operator space into exponentially many invariant subspaces, where the projected Lindbladian acts as an integrable ``Hamiltonian''. We have worked out in detail the case of the quantum ASEP Lindbladian~\cite{Jin2020Stochastic}, and exhibited explicit further examples with arbitrary local dimension. Our study raises several questions. Most prominently, one could wonder what are the consequences of the operator-space fragmentation and integrability beyond correlation functions. For instance, it would be particularly interesting to investigate the dynamics of entanglement-related quantities~\cite{alba2020spreading}. Here, the block diagonal structure of the evolved density matrix gives us a promising starting point for the study of the entanglement negativity~\cite{Vidal2002Computable}. We note that, while using Eq.~\eqref{eq:bound} the latter can be seen to be proportional to $e^{-(J_1+J_2)t/2}$ at late times, its short-time dynamics is expected to be non-trivial. We plan to come back to these questions in future research. 

{\em Acknowledgments:}
We thank Denis Bernard and Aleksandra Ziolkowska for helpful
discussions. This work was supported in part by the EPSRC under grant
EP/S020527/1 (FHLE) and the Alexander von Humboldt foundation (LP).


\appendix

\section{The ASEP Lindbladian}
\label{sec:quantum_asep}

We consider the Lindbladian defined by Eq.~\eqref{eq:liouvillian}. As in the main text, we define the operators $E^{\alpha,\beta}$ by their action $E^{\alpha,\beta}\ket{e^{\gamma}}=\delta_{\beta,\gamma}\ket{e^{\alpha}}$. Explicitly, we have
\begin{align}
	E_{j}^{14}&=\sigma_{j}^{+} \otimes \sigma_{j}^{+}\,,\quad E_{j}^{41}=\sigma_{j}^{-} \otimes \sigma_{j}^{-}\,,\\
	E_{j}^{11}&=\frac{\openone+\sigma_{j}^{z}}{2}\otimes \frac{\openone+\sigma_{j}^{z}}{2}\,,\quad E_{j}^{22}=\frac{\openone-\sigma_{j}^{z}}{2}\otimes \frac{\openone+\sigma_{j}^{z}}{2}\,,\\
	E_{j}^{33}&=\frac{\openone+\sigma_{j}^{z}}{2} \otimes\frac{\openone-\sigma_{j}^{z}}{2}\,,\quad
	E_{j}^{44}=\frac{\openone-\sigma_{j}^{z}}{2}\otimes \frac{\openone-\sigma_{j}^{z}}{2}\,.
\end{align}
With these notations, the Lindbladian superoperator reads
\begin{widetext}
\begin{align} 
	\mathcal{L}&= \sum_{j=1}^{L}\left[ J_{1} E_{j}^{14} E_{j+1}^{41}+J_{2} E_{j}^{41} E_{j+1}^{14}- J_{1} E_{j}^{44} E_{j+1}^{11}-J_{2} E_{j}^{11} E_{j+1}^{44} \right]-\frac{1}{2}\sum_{j=1}^L  \left(J_{1}+J_{2}\right)\left(E_{j}^{22} E_{j+1}^{33}+E_{j}^{33} E_{j+1}^{22}\right)\nonumber\\
	&-\frac{1}{2} \sum_{j=1}^L\left[\left(E_{j}^{22}+E_{j}^{33}\right)\left(J_{1} E_{j+1}^{11}+J_{2} E_{j+1}^{44}\right) + \left(E_{j+1}^{22}+E_{j+1}^{33}\right)\left(J_{2} E_{j}^{11}+J_{1} E_{j}^{44}\right)\right]\,.
	\label{eq:full_asep_lindbladian}
\end{align}
\end{widetext}
We see that $\mathcal{L}$ commutes with $\mathcal{P}^1_j=E_j^{2,2}$, $\mathcal{P}^2_j=E_j^{3,3}$, so that $\mathcal{P}^0_j=\openone_j-(\mathcal{P}^1_j+\mathcal{P}^2_j)$ is the projector onto the subspace associated with density matrices that are diagonal at site $j$. We now proceed to analyze the Lindbladian restricted to the different subspaces $\mathcal{K}_{\boldsymbol{\alpha}}$, defined in the main text.

Let us first consider the case where $\alpha_j=0$, for all $j$, corresponding to the subspace of locally-diagonal density matrices. The Lindbladian restricted to $\mathcal{K}_{\boldsymbol{\alpha}}$ acts on the tensor product of $2$-dimensional local spaces, which are spanned by the states $\ket{e_1}_j$, $\ket{e_4}_j$. Choosing  this basis for all sites, we can rewrite the projected Lindbladian as
\begin{align} 
	\mathcal{L}_{\mathrm{ASEP}}&=\sum_{j=1}^L\left[ J_{1} \sigma_{j}^{+} \sigma_{j+1}^{-}+J_{2} \sigma_{j}^{-} \sigma_{j+1}^{+}\right. \nonumber\\
	-J_{1}&\left.\frac{\left(\openone -\sigma_j^z\right)}{2} \frac{\left(\openone +\sigma_{j+1}^z\right)}{2}-J_{2} \frac{\left(\openone +\sigma_j^z\right) \left(\openone -\sigma_{j+1}^z\right)}{2}\right]\,.
	\label{eq:ASEP_L_periodic}
\end{align}
In the general case, instead, $\alpha_j\neq 0$ for $j\in\{\ell_1,\ldots, \ell_q\}$, and the Lindbladian can be decomposed as in Eq.~\eqref{eq:decomposition}. Suppose first that $|\ell_{k}-\ell_{k+1}|\geq 3$. In this case, it is straightforward to write
\be
\mathcal{L}^{\alpha,\beta}_{[m,n]}=\openone_{m}\otimes\tilde{\mathcal{L}}_{[m+1,n-1]}\otimes \openone_{n}\,,
\ee
where
\begin{align}
	\tilde{\mathcal{L}}_{[m,n]}&=\sum_{j=m}^{n-1}\left[ J_{1} \sigma_{j}^{+} \sigma_{j+1}^{-}+J_{2} \sigma_{j}^{-} \sigma_{j+1}^{+}\right.\nonumber\\
	 -&\left.J_{1}\frac{\left(\openone -\sigma_j^z\right)}{2} \frac{\left(\openone +\sigma_{j+1}^z\right)}{2}-J_{2} \frac{\left(\openone +\sigma_j^z\right) \left(\openone -\sigma_{j+1}^z\right)}{2}\right]\nonumber\\
	&+\frac{(J_2-J_1)}{4}\left(\sigma^z_{m}-\sigma^z_{n}\right)-\frac{(J_1+J_2)}{2}\,.
	\label{eq:ASEP_L_open}
\end{align}
Note that $\tilde{\mathcal{L}}_{[m,n]}$ does not depend on $\alpha$ and $\beta$. As stated in the main text, we can map these Lindbladians onto $XXZ$ Heisenberg Hamiltonians, by considering the similarity transformation
\be
S=\prod_{j=1}^L S_{j}, \qquad S_{j}=\left(\begin{array}{cc}1 & 0 \\ 0 & Q^{j-1}\end{array}\right)_{j}\,,
\label{eq:similarity}
\ee
where $Q=\sqrt{J_{1} / J_{2}}$. Applying this to both Eqs.~\eqref{eq:ASEP_L_periodic} and \eqref{eq:ASEP_L_open}, we finally obtain 
\begin{align}
 S^{-1} \mathcal{L}_{A S E P} S &=\frac{\sqrt{J_{1} J_{2}}}{2} \sum_{j=1}^{L-1} \left[\sigma_{j}^{x} \sigma_{j+1}^{x}+\sigma_{j}^{y} \sigma_{j+1}^{y}\right.\nonumber\\
 +\left.\Delta\left(\sigma_{j}^{z} \sigma_{j+1}^{z}-1\right)\right]&+\frac{\sqrt{J_{1} J_{2}}}{2}\left[2\left(Q^{L} \sigma_{L}^{+} \sigma_{1}^{-}+Q^{-L} \sigma_{L}^{-} \sigma_{1}^{+}\right)\right.\nonumber\\
 +&\left.\Delta\left(\sigma_{L}^{z} \sigma_{1}^{z}-1\right)\right]\,,
\end{align}
and
\be
S^{-1} \tilde{\mathcal{L}}_{[m,n]} S=
\frac{\sqrt{J_{1} J_{2}}}{2}\left\{-H^{XXZ}_{[m,n]}-2\Delta\right\}\,,
\label{eq:XXZ_open}
\ee
where
\be
H^{XXZ}_{[m,n]}=-\sum_{j=m}^{n-1} \left[ \sigma_{j}^{x} \sigma_{j+1}^{x}+\sigma_{j}^{y} \sigma_{j+1}^{y}+\Delta\left(\sigma_{j}^{z} \sigma_{j+1}^{z}-1\right) \right]\,,
\ee
and
\be
\Delta=\frac{Q+Q^{-1}}{2}=\frac{J_{1}+J_{2}}{2\sqrt{J_{1} J_{2}}}\,.
\ee

We are now left to consider the two special cases $\mathcal{L}^{\alpha,\beta}_{[m,m+1]}$ and $\mathcal{L}^{\alpha,\beta}_{[m,m+2]}$. A simple calculation shows that in both cases the restricted Lindbladian is proportional to the identity. Specifically, we find
\begin{align}
\mathcal{L}^{1,1}_{[m,m+1]}&=\mathcal{L}^{2,2}_{[m,m+1]}=0\,,\\ \mathcal{L}^{1,2}_{[m,m+1]}&=\mathcal{L}^{2,1}_{[m,m+1]}=-\frac{J_1+J_2}{2}\,,
\end{align}
and
\be
\mathcal{L}^{\alpha,\beta}_{[m,m+2]}=-\frac{J_1+J_2}{2}\,.
\ee

Finally, let us discuss the coherent term generated by the Hamiltonian~\eqref{eq:hamiltonian}. It is straightforward to rewrite
\begin{align}
	i(H-\bar{H})&=2 i  \sum_{j=1}^L\left[J^{\prime}\left(E_{j}^{11}-E_{j}^{44}\right)\left(E_{j+1}^{22}-E_{j+1}^{33}\right) \right.\nonumber\\
	+\left(E_{j}^{22}\right.&-\left.\left.E_{j}^{33}\right)\left(E_{j+1}^{11}-E_{j+1}^{44}\right)\! +\! h \left(E_{j}^{22}-E_{j}^{33}\right)\right]\,.
\end{align}
From this expression, it follows immediately that the projection of this operator onto each invariant subspace results in diagonal boundary terms, thus preserving integrability. 

\section{Higher spin generalizations}
\label{sec:higher_spin}

We consider a Hilbert space $\mathcal{H}=\mathcal{H}_1\otimes \cdots \otimes \mathcal{H}_L$, with $\mathcal{H}_j\simeq \mathbb{C}^{N}$ and a purely dissipative Lindblad equation of the form $\dot{\rho}=-\mathcal{D}(\rho)$, with
\begin{align}
\mathcal{D}(\rho)=\sum_{j=1}^L\sum_{\alpha,\beta=1}^N \left[ L^{(\alpha,\beta)}_{j}\rho(t) L^{(\alpha,\beta)\dagger}_{j}\right.\nonumber\\
-\left.\frac{1}{2}\left\{L^{(\alpha,\beta)\dagger}_{j}L^{(\alpha,\beta)}_{j},\rho(t)\right\}\right]\,,
\end{align}
where
\be
L^{(\alpha,\beta)}_{j}=E^{\alpha,\beta}_{j}E^{\beta,\alpha}_{j+1}\,.
\ee
Here, $E^{\alpha,\beta}$ is the operator with matrix elements $(E^{\alpha,\beta})_{\alpha^\prime,\beta^\prime}=\delta_{\alpha,\alpha^\prime}\delta_{\beta,\beta^\prime}$. The Lindbladian superoperator reads
\begin{widetext}
\begin{align}
	\mathcal{L}&=\sum_{j=1}^L\sum_{\alpha,\beta=1}^N \left[\left(E^{\alpha,\beta}_{j}\otimes E^{\alpha,\beta}_{j}\right)\left(E^{\beta,\alpha}_{j+1}\otimes E^{\beta,\alpha}_{j+1}\right)
	-\frac{1}{2}\left(E^{\beta,\beta}_{j}E^{\alpha,\alpha}_{j+1}\otimes \openone+\openone\otimes E^{\alpha,\alpha}_{j}E^{\beta,\beta}_{j+1}\right)\right]\nonumber\\
&=\sum_{j=1}^L\left[\sum_{\alpha,\beta=1}^N \left(E^{\alpha,\beta}_{j}\otimes E^{\alpha,\beta}_{j}\right)\left(E^{\beta,\alpha}_{j+1}\otimes E^{\beta,\alpha}_{j+1}\right)
-\openone\right]=: \mathcal{L}^\prime-L \openone\,,
\end{align}
\end{widetext}
where we used $\sum_{\alpha=1}^N E_j^{\alpha,\alpha}=\openone$. In the superoperator formalism, it is readily seen that the Lindbladian
$\mathcal{L}^\prime$ (and hence $\mathcal{L}$) commutes with the $d(d-1)$ rank-$1$ projectors
\be
\mathcal{P}^{(\alpha,\beta)}_j= E^{\alpha,\alpha}_j\otimes
E^{\beta,\beta}_j\,,\qquad \alpha\neq \beta,\quad  \alpha,\beta=1,\ldots d\,,
\ee
so that 
\be
\mathcal{P}^{0}_j=\openone_j-\sum_{\alpha\neq\beta}\mathcal{P}^{(\alpha,\beta)}_j=\sum_{\alpha=1}^N E^{\alpha,\alpha}_j\otimes E^{\alpha,\alpha}_j\,,
\ee
is a projector onto the subspace spanned by the states
\be
\ket{e_\alpha}=\ket{\alpha}\otimes \ket{\alpha}\rangle\,,\qquad \alpha=1\,,\ldots \,, N\,.
\label{eq:diagonal_states}
\ee
This implies that the Hilbert space $\mathcal{H}\otimes \mathcal{H}$ splits into exponentially many invariant subspaces $\mathcal{K}_{\boldsymbol{\alpha}}$.

We start by analyzing the subspace corresponding to $\alpha_j=0$ for all $j$. First, it is straightforward to show that the local terms in the Lindbladian $\mathcal{L}^\prime$ act as swap operators, when applied to the states~\eqref{eq:diagonal_states}. Explicitly, we have
\begin{align}
\sum_{\alpha,\beta=1}^N \left(E^{\alpha,\beta}_{j}\otimes E^{\alpha,\beta}_{j}\right)\left(E^{\beta,\alpha}_{j+1}\otimes E^{\beta,\alpha}_{j+1}\right)\ket{e_\gamma}_{j} \ket{e_\delta}_{j+1}\nonumber\\
=\ket{e_\delta}_{j}\otimes \ket{e_\gamma}_{j}=\Pi_{j,j+1}\ket{e_\gamma}_{j} \ket{e_\delta}_{j+1}\,,
\label{eq:permutation_explicit}
\end{align}
where $\Pi_{j,j+1}$ is the operator permuting sites $j$ and $j+1$.  Thus, we simply obtain
\be
\mathcal{L}^\prime\big|_{\boldsymbol{\alpha}=0}=\sum_{j=1}^{L}\Pi_{j, j+1}\,,
\ee
which is Eq.~\eqref{eq:final_lind_N} in the main text. Next,  suppose $\alpha_j\neq 0$ for $j\in\{\ell_1,\ldots, \ell_q\}$. Then, the Lindbladian $\mathcal{L}^\prime$ can be decomposed as in Eq.~\eqref{eq:decomposition}. First, we consider $|\ell_{k}-\ell_{k+1}|\geq 3$. In this case, we see that the factorization condition~\eqref{eq:factorization} holds, with
\begin{align}
	\tilde{\mathcal{L}}_{[m,n]}=\sum_{j=m}^{n-1}\left[\sum_{\alpha,\beta=1}^N \left(E^{\alpha,\beta}_{j}\otimes E^{\alpha,\beta}_{j}\right)\left(E^{\beta,\alpha}_{j+1}\otimes E^{\beta,\alpha}_{j+1}\right)\right]\,.
\end{align}
Choosing the diagonal basis~\eqref{eq:diagonal_states}, and making use of Eq.~\eqref{eq:permutation_explicit}, we obtain Eq.~\eqref{eq:open_suN}. Finally,  when $|\ell_j-\ell_{j+1}|\leq 3$, it is easy to show that the projection of $\mathcal{L}^\prime$ is vanishing.

\bibliography{./bibliography}

\end{document}